\documentclass[aps,pre,twocolumn,superscriptaddress,showpacs,showkeys]{revtex4-2}
\usepackage{graphicx}                   
\usepackage{hyperref}                   
\usepackage{amsmath,amssymb}            
\usepackage{epstopdf}
\usepackage{subfigure}					
\usepackage{float}
\usepackage{color}
\definecolor{orange}{rgb}{1,0.5,0}
\definecolor{brown}{rgb}{0.65, 0.16, 0.16}
\definecolor{phlox}{rgb}{0.87, 0.0, 1.0}

\graphicspath{{figs/}}					

\begin{document}

	\title{Sandpiles with finite-range interactions}

	\author{Abbas Shoja-Daliklidash}
	\affiliation{Department of science, University of Mohaghegh Ardabili, Ardabil, Iran}
    \author{M. Nattagh-Najafi}
	\affiliation{Department of science, University of Mohaghegh Ardabili, Ardabil, Iran}

    \email{morteza.nattagh@gmail.com}

    \begin{abstract}
    We investigate the sandpile model with Yukawa-type interactions, whose effective range is tuned by an external parameter $R$. Our results reveal that at specific values of $R$, the system exhibits giant avalanches that span the system, leading to percolation. The probability of such giant avalanches demonstrates two distinct regimes as a function of $R$: for sufficiently small $R$, it increases monotonically, whereas for large $R$ it undergoes threshold dynamics, so that at certain values of $R$, the percolation probability exhibits abrupt jumps. We refer it to as \textit{pseudo-percolation transitions}, based on which we propose a hierarchical percolation model at the mean-field level: each percolation transition corresponds to percolation within a disc of radius $R$. We further examine both local and global geometrical observables. The local quantities include avalanche size, mass, and duration and sub-avalanche mass, while for the global characterization we analyze the loop length and gyration radius of the external perimeter, as well as the mass of sub-avalanches. Remarkably, all these observables exhibit power-law scaling for all values of $R$, with exponents that vary systematically with $R$. Notably, in the vicinity of the pseudo-percolation transition points, the exponents approach characteristic values, signaling a distinct critical behavior. 
    \end{abstract}

	\maketitle
	
\section{Introduction}

Self-organized criticality (SOC) refers to the emergence of critical behavior in dynamical systems without the necessity of fine-tuning external control parameters. Since its introduction, SOC has been identified across a remarkably broad spectrum of natural and artificial systems. Examples include neural dynamics in the brain~\cite{beggs2003neuronal,de2006self,ribeiro2010spike,hesse2014self,priesemann2014spike,plenz2021self}, rough surfaces and roughening processes~\cite{kondev2000nonlinear}, liquid foams~\cite{ritacco2020complexity}, sheared suspensions~\cite{corte2009self}, atmospheric cascades~\cite{rybczynski2001self,yano2012self}, earthquakes~\cite{bak1989earthquakes,sornette1989self,ito1990earthquakes,chen1991self,rahimi2022self,najafi2020avalanches}, stock markets~\cite{stauffer1999self,biondo2015modeling,bartolozzi2005self,aleksiejuk2002self,dupoyet2011replicating}, social systems~\cite{batty1999self,weisbuch2001social,brunk2002societies,kron2009society}, astrophysical processes~\cite{aschwanden2012self}, knowledge creation~\cite{tadic2017mechanisms}, water diffusion in porous media~\cite{najafi2016water}, rainfall patterns~\cite{andrade1998analysis}, electronic avalanches in electron gases~\cite{najafi2018percolation,najafi2019electronic,najafi2021flicker}, vortex avalanches in superconductors~\cite{vlasko2004experimental,wijngaarden2006avalanches}, cumulus cloud dynamics~\cite{najafi2021self,cheraghalizadeh2022simulating}, real sandpiles and rice piles~\cite{aegerter2003avalanche}, and self-organized Lévy flights~\cite{boguna1997long}, among many others. For a comprehensive overview of SOC phenomena, we refer the reader to~\cite{dhar1999some,najafi2021some}. In this work, we contribute to this field by investigating a long-range sandpile model, which provides new insights into the interplay between interaction range and critical avalanche dynamics.\\

The influence of long-range interactions on critical phenomena has long been a fundamental and intriguing question. Such interactions can modify both the universality class and the critical exponents of a system, with the degree of modification depending on their strength and decay properties, as discussed by Zeng et al.~\cite{zeng2022theory}. For instance, when interactions decay according to a power-law relationship with distance, the effective dimensionality of the system is altered, and the nature of this alteration depends on the decay exponent~\cite{fisher1972critical,aizenman1988critical}. Consequently, scaling laws and phase transitions may deviate significantly from those predicted by the conventional theory of short-range critical phenomena. Using renormalization group arguments, Fisher~\cite{fisher1972critical} demonstrated that for an isotropic $n$-component order parameter with long-range attractive interactions decaying as $1/r^{\sigma+d}$ (where $d$ denotes the spatial dimension), the critical behavior is governed by the parameter $\epsilon \equiv 2\sigma - d$. Specifically, when $\epsilon < 0$, the system exhibits classical (mean-field) critical exponents; when $\sigma > 2$, the exponents coincide with those of short-range interactions. At the marginal case $\epsilon = 0$, the scaling relations acquire logarithmic corrections involving fractional powers of the tuning parameter.\\

In this work, we address the same problem in the context of sandpiles for $d=2$ and $\sigma = -1$. Traditionally, the influence of long-range interactions in sandpile models has been investigated through the introduction of long-range links in complex networks, such as random networks~\cite{najafi2014bak}, scale-free networks~\cite{goh2003sandpile,lee2004sandpile}, and regular lattices augmented with long-range connections, such as small-world networks~\cite{hoore2013critical,najafi2018statistical}; see also~\cite{najafi2021some}. In these systems, a crossover between the behavior of random graphs and regular lattices is typically observed~\cite{hoore2013critical}. Long-range interactions are particularly intriguing because they allow particles to move locally while being influenced by nonlocal forces, which in turn strongly affects the universality class of the sandpile. From a mathematical standpoint, the thermodynamics of sandpile systems—viewed as out-of-equilibrium systems with long-range interactions—remains an open problem~\cite{dauxois2002dynamics,bouchet2010thermodynamics}. It has not been systematically analyzed to date. Importantly, the scaling properties of the model can change dramatically depending on both the nature and the range of the interactions. For example, when interactions are divisible, meaning that each toppling grain can split into an arbitrary number of smaller grains, the limiting field may correspond either to a fractional Gaussian field or to a bi-Laplacian field, depending on the decay exponent of the interactions~\cite{chiarini2021constructing}. In the present paper, we demonstrate that a hierarchy of pseudo-percolation transitions emerges, which significantly influences the thermodynamic properties of the system. Near each transition point, the critical exponents undergo successive changes in their values.\\

The remainder of the paper is organized as follows. In the next section, we discuss the thermodynamic features of systems with long-range interactions. The details of the model are presented in Section~\ref{SEC:model}, while the results of numerical simulations are reported in Section~\ref{SEC:sim}. Section~\ref{SEC:fractal} is devoted to the analysis of the fractal properties of the model. Finally, we conclude with a summary of the main findings and their implications.

\section{Thermodynamic Limit of Long-Range Interacting Systems: \\
A Dimensional Analysis}

In statistical mechanics, the existence of a well-defined thermodynamic limit is a cornerstone for connecting microscopic dynamics with macroscopic observables. For systems with short-range interactions this limit is guaranteed: the total interaction energy scales extensively with the system size, ensuring that the energy per particle or per unit volume remains finite as $N,V \to \infty$. However, in systems governed by long-range interactions, such as Coulomb or gravitational forces, extensivity can break down. In such cases, the interaction energy grows faster than the system volume, causing divergences in thermodynamic quantities. Understanding how the spatial dimension $d$ competes with the interaction exponent $\sigma$ therefore becomes essential for characterizing the critical behavior of such systems.

To illustrate this point, let us begin with the Coulomb interaction
\begin{equation}
	U(r)=\frac{u}{r},
\end{equation}
where $u$ is a constant. The self-energy of a sphere of radius $R$ is found to be
\begin{equation}
	\begin{split}
		E(R)&=\int_0^R\left(\frac{4}{3}\pi r^3\right)U(r)(4\pi r^2 \rho dr)\\
		&=\frac{4\pi u}{5}\rho^2R^2V(R),	
	\end{split}
\end{equation}
where $V(R)$ is the sphere volume. This result shows that Coulomb (or gravitational) interactions are too long-ranged to permit a well-defined thermodynamic limit, since $E(R)/V(R)\to \infty$ as $R\to\infty$.  

To generalize the argument, we consider the following interaction in $d$ dimensions: 
\begin{equation}
	U_{\sigma}(r)\equiv \frac{u}{r^{d+\sigma}}, 
\end{equation}
where $\sigma$ is a tunable exponent controlling the decay of the potential. The total energy is then
\begin{equation}
	E(R)=\frac{1}{2}\int_{\Omega}d^drd^dr'\rho(\textbf{r})\rho(\textbf{r}')U_{\sigma}(\textbf{r}-\textbf{r}'),
    \label{eq: 4}
\end{equation}
where $\rho$ is the particle density and $\Omega$ the system volume. For a uniform density $\rho(\textbf{r})=\rho$, the energy per unit volume reads
\begin{equation}
	e(R)\equiv \frac{E(R)}{V(R)}= \frac{u\rho^2S_dR^{-\sigma}}{2\sigma}\left[(a/R)^{-\sigma}-1\right],
\end{equation}
where $S_d$ is the surface area of the unit sphere, and $a$ a UV cutoff (e.g. a lattice spacing or particle hard-core radius). This integral is convergent only when $\sigma>0$. For $\sigma=0$ one obtains
\begin{equation}
	\lim_{\sigma\to 0}e(R)= \frac{u\rho^2S_d}{2}\log(R/a),
\end{equation}
which diverges logarithmically as $R\to\infty$, signaling an infrared (IR) divergence. Within renormalization group frameworks, such logarithmic divergences may sometimes be absorbed into the redefinition of physical parameters, but strong divergences for $\sigma<0$ indicate a fundamental breakdown of extensivity.

A more refined analysis requires allowing density fluctuations,
\begin{equation}
	\rho(\textbf{r})=\rho+\delta \rho(\textbf{r}),
\end{equation}
where 
\begin{equation}
	\left\langle \delta \rho(\textbf{r})\delta \rho(\textbf{r}')\right\rangle =\rho_0^2 |\textbf{r}-\textbf{r}'|^{-\beta}, \ \ \left\langle \delta \rho(\textbf{r})\right\rangle =0,
\end{equation}
with $\beta$ an exponent characterizing spatial correlations of density fluctuations. When $\rho\ne 0$, this correction gives a subleading contribution, whereas for $\rho=0$ we find 
\begin{equation}
	e(R)= \frac{u\rho_0^2S_dR^{-\sigma-\beta}}{2(-\sigma-\beta)}\left[1-(a/R)^{-\sigma-\beta}\right],
\end{equation}
which remains convergent provided that $\sigma+\beta>0$.  

These considerations link directly to the renormalization group analysis of Fisher and co-workers~\cite{fisher1972critical}, who demonstrated that for an $n$-component order parameter the critical behavior depends sensitively on $\sigma$: when $\sigma<d/2$ the critical exponents retain their classical (mean-field) values, whereas for $\sigma>2$ they coincide with those of short-range interactions. The marginal case $\sigma=d/2$ exhibits logarithmic corrections, where pure power-law scaling is modulated by logarithmic terms.  

In the present study, we focus on the case $\sigma=-1$ for two-dimensional sandpiles. This interaction is so long-ranged that severe IR divergences emerge. To regularize the problem, we introduce a Yukawa-type cutoff, motivated physically by screening effects (e.g., Debye or Thomas–Fermi screening in plasmas and electronic systems). The Coulomb potential is thus replaced by
\begin{equation}
	U(r)\to U_{\text{Yukawa}}(r)\equiv \frac{ue^{-r/R}}{r},	
\end{equation}
where $R$ plays the role of a screening length, effectively acting as an IR cutoff.  

From the perspective of sandpile dynamics, this modification has a clear physical meaning: while each toppling event can in principle affect sites at arbitrarily large distances, the Yukawa cutoff ensures that this influence decays exponentially beyond $R$. Consequently, the system interpolates between two regimes: for scales much smaller than $R$, the dynamics resemble those of a pure long-range sandpile, while at scales larger than $R$ the system behaves effectively as a short-range model. As we will show in later sections, this interpolation profoundly affects the universality class, giving rise to pseudo-percolation transitions and a hierarchy of scaling behaviors.

\section{The model}\label{SEC:model}

We construct our sandpile model on a two–dimensional square lattice of size $L \times L$. To each site $i$ we assign an integer variable $n_i$ denoting the number of sand grains located at that site~\cite{najafi2016bak}. In contrast to the conventional BTW model, here every sand grain carries an electric charge $\delta q=+1$, while each site has a background charge $Q_0=-4$. The total charge at site $i$ is therefore
\begin{equation}
q_i=n_i\delta q+Q_0=n_i-4,
\end{equation}
so that $n_i$ may also be interpreted as the effective “height” variable of the sandpile.

\subsection{Energy Functional and Long–Range Coupling}
The energetic cost of a given configuration is defined in terms of a local self–energy together with a Yukawa–type long–range interaction:
\begin{equation}
U\left\lbrace q\right\rbrace =\frac{1}{2}\sum_{i=1}^{N}q_i^2
+\frac{\alpha}{2R}\sum_{i\neq j}^{N}q_iq_j\,
\frac{\theta_{R}(r_{ij})\,e^{-\frac{r_{ij}}{R}}}{r_{ij}},
\end{equation}
where $r_{ij}$ is the distance between sites $i$ and $j$, $\alpha$ is an external parameter (coupling constan), and $R$ is the Yukawa cutoff length. The step function
\begin{equation}
\theta_{R}\left(r_{ij}\right)=
\begin{cases}
0 & r_{ij}>R \ \text{or}\ r_{ij}=0,\\
1 & \text{otherwise},
\end{cases}
\end{equation}
ensures that only sites within a disk of radius $R$ contribute to the interaction. This construction interpolates smoothly between two limiting regimes: for distances $r\ll R$, the interaction is essentially Coulombic ($\sim 1/r$), whereas for $r\gg R$ the exponential factor suppresses the coupling, yielding effectively local dynamics.

At the mean–field level, one can estimate the average contribution of the long–range term as
\begin{equation}
U \propto A\frac{\bar{\rho}^2\alpha}{2R}\int d\mathbf{r}\,
\left(\Theta_{R}(r) r^{-1}e^{-r/R}\right)
=A\widetilde{\alpha}\bar{\rho}^2,
\end{equation}
with $\widetilde{\alpha}=\pi\alpha\left(1-e^{-1}\right)$, $A=L^2$ the system area, and $\bar{\rho}=\sum_{i} q_i/A$ the mean charge density. The explicit $1/R$ factor in front regularizes the infrared divergence that would otherwise arise for purely Coulombic interactions as $R\to\infty$.

\subsection{Instability Criterion and Toppling Rule}

\begin{figure*}
	\centering
	\includegraphics[width=0.6\linewidth]{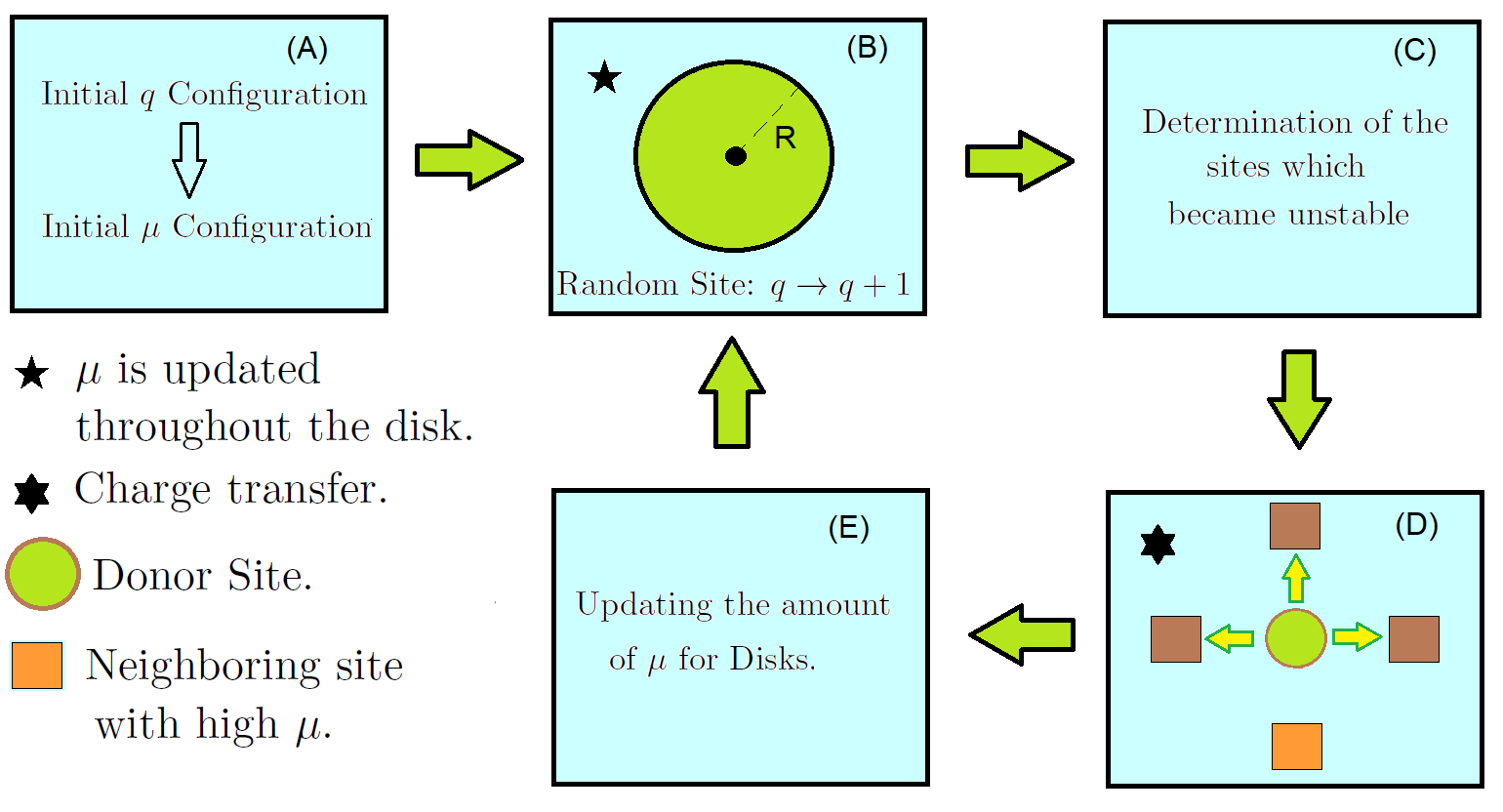}
	\caption{Schematic graph of the toppling algorithm for the dynamics of the long-range interaction sandpile model.  }%
	\label{fig:algorithm}	
\end{figure*}

\begin{figure}
	\centering
	\includegraphics{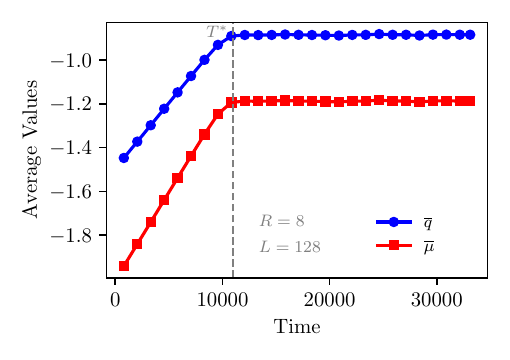}
	\caption{Time evolution of $q$ and $\mu$ for $L=128$, showing the transient regime and the onset of the stationary state. }%
	\label{fig:Stationary}	
\end{figure}

The local chemical potential is defined as the functional derivative of the energy with respect to $n_i$:
\begin{equation}
\mu_i=\frac{\delta U}{\delta n_i}
=q_i+\frac{\alpha}{R}\sum_{j\neq i}^{N}{q_j\,
\frac{\theta_{R}(r_{ij})e^{-\frac{r_{ij}}{R}}}{r_{ij}}}.
\end{equation}
A site $i$ becomes unstable whenever $\mu_i>0$. In that case, the site topples by losing
\begin{equation}
n_p(i)=\min\{q_i+3,\,4\}
\end{equation}
grains. These grains are then redistributed among the neighbors. To determine the receivers, we sort the four nearest neighbors according to their chemical potentials,
\begin{equation}
\{s_j\}_{j=1}^{n_p(i)} \equiv 
\text{sort}\{\text{neighbors}_j\}_{j=1}^4,
\end{equation}
and transfer grains in ascending order. In this way, neighbors with smaller $\mu$ have a higher probability of receiving particles. If the site lies at the boundary of the lattice, one or more grains may leave the system, thereby providing the dissipation mechanism. To treat boundaries consistently, we assign virtual chemical potentials $\mu_{\text{Left}},\mu_{\text{Right}},\mu_{\text{Down}},\mu_{\text{Up}}$ to the four edges, such that grains can dissipate preferentially through the boundary with the lowest effective occupation.

\subsection{Simulation Procedure}
The algorithmic implementation is illustrated in Fig.~\ref{fig:algorithm}. The system is initialized with a random charge configuration, ensuring that all $q_i$ values lie within $[-3,0]$, which corresponds to stable states. The simulation proceeds as follows:  
1. Compute the chemical potential array $\{\mu_i\}$ from Eq.\ref{eq: 4}.  
2. Add one unit of charge to a randomly chosen site $i_0$, i.e. $q_{i_0}\to q_{i_0}+1$.  
3. If $\mu_{i_0}>0$, this site and possibly others within the Yukawa interaction disk $D(i_0,R)$ may become unstable. A stack data structure (arrays \texttt{stackx}, \texttt{stacky}) keeps track of all unstable sites.  
4. Iteratively topple unstable sites: each toppling releases $n_p(i)$ grains, distributed according to the neighbor–sorting rule or dissipated at the boundaries.  
5. Update the local charges and recompute chemical potentials for affected neighborhoods. New unstable sites are pushed onto the stack until the avalanche terminates. This procedure defines a driven dissipative dynamics that, as will be shown in subsequent sections, exhibits crossover behavior between long–range and effectively local universality classes, depending on the value of the cutoff length $R$.

To investigate the dynamical behavior of the model, we perform large-scale simulations on a square lattice of size $L\times L$. The system is initialized from a random configuration of charges $q_i$ drawn uniformly within the interval $[-3,0]$. At each time step, a single grain (unit charge) is added at a randomly chosen site $i_0$, which leads to an update of the local chemical potential profile. If this perturbation renders the site unstable ($\mu_{i_0}>0$), the site topples following the rule defined in Section~\ref{SEC:model}. This local instability can propagate to neighboring or even distant sites, thereby generating a cascade of topplings. We refer to such cascades as \emph{avalanches}, which begin with the injection of a grain and terminate once no unstable sites remain in the system.

We define the external ``time" variable $t$ as the total number of grains injected into the lattice. As in standard sandpile models~\cite{najafi2021some}, the dynamics separate into two regimes: a transient regime, which is only visited once and depends on the initial condition, and a recurrent (stationary) regime, in which the system self-organizes and statistical observables attain stationary distributions. Figure~\ref{fig:Stationary} illustrates this behavior for $L=128$ and $R=8$: after a crossover time $T^*$, both the spatially averaged charge $\bar{q}(t)$ and the spatially averaged chemical potential $\bar{\mu}(t)$ fluctuate around well-defined steady values. All subsequent measurements reported in this work are performed in this stationary regime.
\subsection{The Statistical Quantities of Interest}
Generally, an avalanche is composed of many connected components, called \textit{sub-avalanches}. Our analysis focuses on two classes of observables: local and global. We analyze local and global properties of sub-avalanches, as well as the avalanches as the whole. 

The local observables are listed below:
   \begin{itemize}
      \item Avalanche size $s$: total number of topplings within an avalanche.
      \item Avalanche mass $m$: number of distinct sites that topple at least once.
       \item Sub-mass $sm$: number of distinct sites involved in a sub-avalanche.
      \item Avalanche duration $d$: number of parallel update steps required until the avalanche ceases.
   \end{itemize}

The global (geometric) observables are listed as follows:
   \begin{itemize}
      \item Loop length $l$: length of the external perimeter of a sub-avalanche cluster.
      \item Gyration radius $r$: radius of gyration of the cluster’s external boundary.
   \end{itemize}
The scaling relations of these quantities define some scaling exponents, which identify the universality class of the system.
In addition, we monitor the percolation probability (PP), defined as the fraction of sub-avalanches that connect one side of the system to the opposite boundary. This observable serves as a probe of pseudo-percolation transitions in the system.

For all of these quantities $x\in\{s,m,d,sm,l,r\}$, we find that the probability distribution obeys the finite-size scaling form
\begin{equation}
P(x)\propto x^{-\tau_x}F\left(x,L\right) ,
\end{equation}
where $\tau_x$ is the scaling exponent, and $F$ is a universal finite size scaling function which gives the scaling relation between $x$ and $L$. Furthermore, different observables are related by algebraic scaling relations of the form
\begin{equation}
y\propto x^{\gamma_{yx}},
\end{equation}
with the cross-exponents $\gamma_{xy}$ satisfying the hyperscaling relation
\begin{equation}
\gamma_{yx}=\frac{\tau_y-1}{\tau_x-1}.
\label{Eq:gamma}
\end{equation}
For comparison, in the ordinary BTW sandpile model the fractal dimension of avalanche clusters is known to be $\gamma_{rl}\equiv D_f^{\text{BTW}}=5/4$~\cite{lubeck1997large}. A comprehensive review of the BTW critical exponents can be found in~\cite{dhar1999some,najafi2021some}.

\section{Hierarchical Pseudo-Percolation Transitions, A HMF Description}
\begin{figure}
	\centering
	\includegraphics{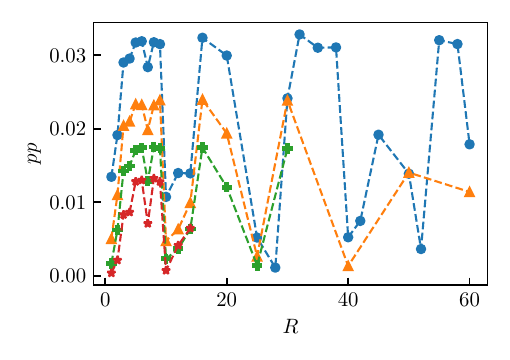}
	\caption{PP in terms of $R$ for $L=64, 128, 256$ and $512$ (from top to bottom). }%
	\label{fig:pp}	
\end{figure}

In this section we develop a hierarchical mean field (HMF) theory for the model. To this end, we first turn our attention to the PP, which helps us to understand the global statistical properties of the model. PP serves as a fundamental global observable and forms the starting point of our analysis. The PP quantifies the likelihood that a large-scale avalanche spans the system by connecting two opposite boundaries. This quantity is particularly significant because it bridges the behavior of local observables, with global structural properties of the model.\\

In Figure~\ref{fig:pp}, we present the PP for different system sizes $L=64, 128, 256,$ and $512$ as a function of $R$. As is well known in sandpile models~\cite{najafi2021some}, PP decreases systematically with increasing system size $L$, and the corresponding curves shift upward almost rigidly as $L$ decreases. This behavior is consistent with that observed in the ordinary BTW model. One may initially regard $R$ as the effective range of avalanches, an assumption that proves only partially valid in the small-$R$ regime, as supported by both simulations and the mean-field argument. Specifically, for small $R$, PP is an increasing function of $R$, reflecting the fact that a larger interaction range allows more sites to be affected by a local perturbation. Consequently, avalanche frontiers expand more rapidly and smoothly, and the effective avalanche range increases with $R$. However, for sufficiently large $R$, a competing mechanism emerges: an avalanche initiated at site $i$ may trigger activity at a distant site $j$, but this activity can become disconnected from the original cluster, effectively splitting the avalanche into disjoint parts, each with smaller spatial extent. This leads to a reduction in PP, as illustrated in Figure~\ref{fig:pp}. Moreover, for large $R$, oscillations with nearly constant frequency across all $L$ values are observed.\\

To theoretically describe the PP, we first consider the BTW model, which lies in the diffusion regime. In this case,
\begin{equation}
    \left\langle r_{\text{avalanche}}\right\rangle \propto t^{1/2},
\end{equation}
where $r_{\text{avalanche}}$ is the average avalanche radius, and $t$ denotes the \textit{internal time}, i.e., the duration between the initiation and termination of an avalanche. One may interpret a grain as a random walker that starts from the lattice center and diffuses until it reaches the boundaries. The number of random walks required to explore a region of linear scale $\zeta$ is
\begin{equation}
    n_{\text{rw}}(\zeta) \propto \zeta^{\chi},
\end{equation}
with $\chi=2$ for normal diffusion. On the other hand, the number of topplings occurring during this process (denoted by $s_{\zeta}$) is proportional to $n_{\text{rw}}(\zeta)$, yielding
\begin{equation}
    s_{\zeta} \propto \zeta^{\chi}.
\end{equation}
It is well established that the PP of avalanches in sandpile models scales as~\cite{najafi2021some}
\begin{equation}
    P(L) \propto L^{-\gamma},
    \label{Eq:percolation}
\end{equation}
where $L$ is the system size.

\begin{figure}[h]
    \centering
    \includegraphics[width=0.8\linewidth]{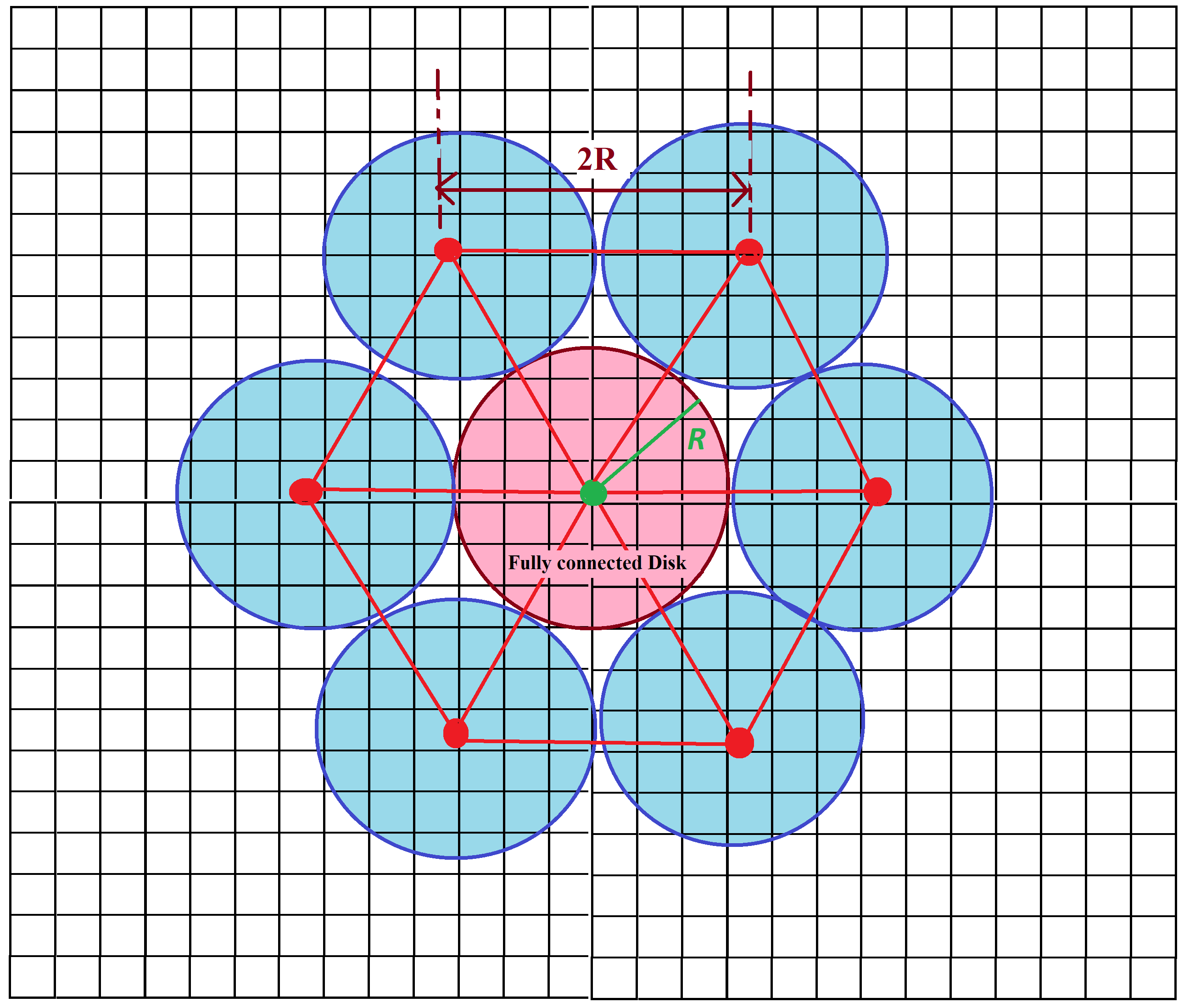}
    \caption{Schematic representation of dividing the lattice into blocks of radius $R$.}
    \label{fig:Schematic}    
\end{figure}

To analyze our model, we subdivide the lattice into blocks of linear size $R$, as illustrated in Fig.~\ref{fig:Schematic}. This approach enables us to distinguish between the small-scale dynamics ($\ll R$) and the large-scale dynamics ($\gg R$). For large scales, we assume particles perform diffusive motion, whereas the small-scale regime exhibits qualitatively different dynamics. Such a subdivision can also be applied to the ordinary sandpile model with arbitrary $R$, for which Eq.~\ref{Eq:percolation} can be rewritten as
\begin{equation}
    P(L) \propto (L/R)^{-\gamma}R^{-\gamma} = (L/R)^{-\gamma} s_R^{-\gamma/\chi},
\end{equation}
which is manifestly independent of $R$. Equivalently,
\begin{equation}
    \begin{split}
        P(\text{giant avalanche}) &= P(\text{perc. inside disks}) \\
        & \times P(\text{perc. outside}),
    \end{split}
\end{equation}
with
\begin{equation}
    \begin{split}
        & P(\text{perc. outside}) \propto (L/R)^{-\gamma}, \\
        & P(\text{perc. inside disks}) \propto s_R^{-\gamma/\chi}.
    \end{split}
\end{equation}

For the long-range sandpile model, We take a same strategy: the percolation should occur both inside and outside the discs simultaneously. One expects that the percolation outside (between) the discs be the same as the ordinary BTW model, this time for the reduced linear size $L/R$. Inside the discs the situation is different given that there is almost all-to-all interaction between the sites. Let us denote the average range of avalanches inside the discs by $R_1<R$. One can find this average radius using a dynamical argument: the avalanche range $R_1$ grows until the point at which all the sites inside it become stable and the average chemical potential and the average charge at a typical point $x$, $\bar{\mu}_{R_1}(\textbf{x})$ and $\bar{q}_{R_1}(\textbf{x})$ become equal the the global averages. This means that $\bar{\mu}_{R_1}(\textbf{x})$ and $\bar{q}_{R_1}(\textbf{x})$ in the following relation: 
\begin{equation}
\bar{\mu}_{R_1}(\textbf{x})=\bar{q}_{R_1}(\textbf{x})+\frac{\alpha}{R}\int_{|\textbf{y}-\textbf{x}|\le R_1}\frac{q(\textbf{y})e^{-|\textbf{y}-\textbf{x}|/R}}{|\textbf{y}-\textbf{x}|}\text{d}^2\textbf{y},
\end{equation}
are replaced by $\bar{\mu}(R)$ and $\bar{q}(R)$ which are the global averages of $\mu$ and $q$ for a system with parameter $R$, so that
\begin{equation}
\bar{\mu}(R)=\bar{q}(R)+\frac{\alpha}{R}\int_0^{R_1}\frac{q(r)e^{-r/R}}{r}\text{d}^2\textbf{r}.
\label{eq:MFmu}
\end{equation}
$R_1$ is referred to as the \textit{effective block radius}. We then adopt a similar strategy t the BTW case with distinct critical exponents. We hypothesize that
\begin{equation}
    \begin{split}
        & P(\text{perc. outside}) \propto (L/R)^{-\gamma_1}, \\
        & P(\text{perc. inside disks}) \propto s_{R_1}^{-\xi},
    \end{split}
\end{equation}
where $\gamma_1$, $\gamma_2$, and $\xi \equiv \gamma_2/\chi$ are scaling exponents, with $\gamma_2$ not necessarily equal to $\gamma_1$. In the HMF level, we replace $q(r)$ in equation~\ref{eq:MFmu} with $\bar{q}(R)$, leading to
\begin{equation}
    R_1 = -R \ln\left[1-f(R)\right],
    \label{Eq:R1}
\end{equation}
with
\begin{equation}
    f(R) \equiv \frac{\bar{q}(R) - \bar{\mu}(R)}{2\pi \alpha |\bar{q}(R)|}.
\end{equation}

\begin{figure}[h]
    \centering
    \includegraphics{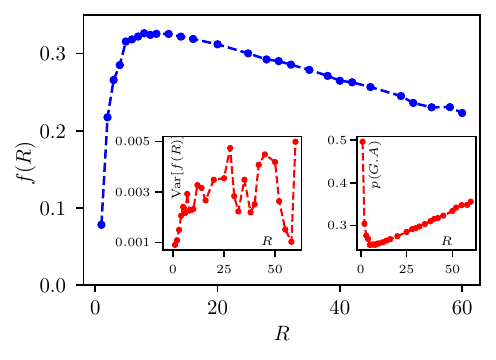}
    \caption{Main plot: behavior of $f(R)$ as a function of $R$ for $L=64$, with $\gamma=0.5$, $\gamma_1=0.45$, and $\alpha=0.1$. The right inset shows the giant avalanche probability $P(\text{G.A.})$ as a function of $R$, while the left inset displays the variance of $f(R)$, highlighting strong fluctuations for $R>10$.}
    \label{fig:f_R}    
\end{figure}

\begin{figure}[h]
    \centering
    \includegraphics[width=0.7\linewidth]{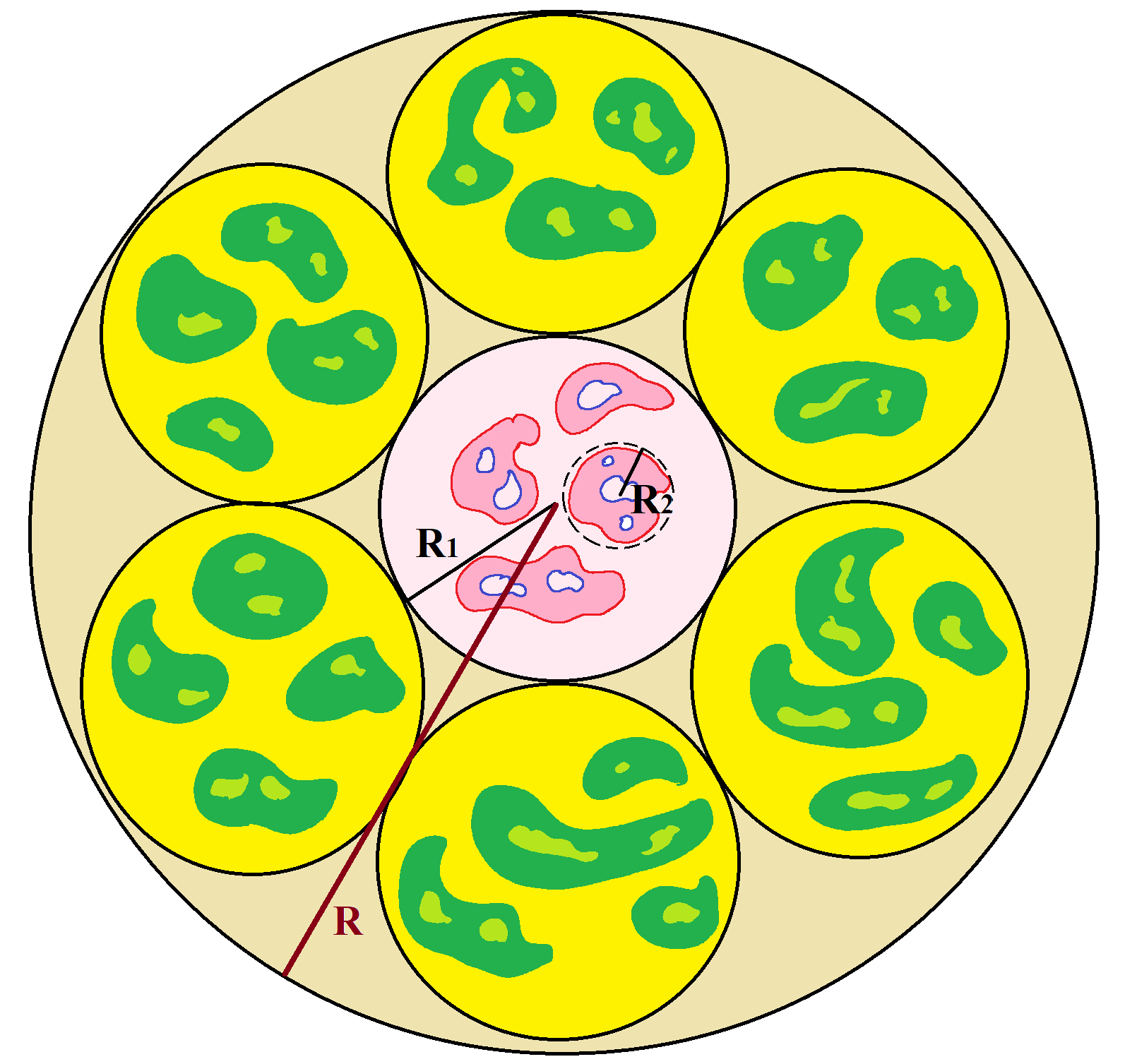}
    \caption{Hierarchical PP in the HMF theory. The large disc denotes the original block ($R$), while smaller discs represent successive hierarchical subdivisions: $R_1$, $R_2$, etc. Colored regions indicate avalanches.}
    \label{fig:Hierarchical}    
\end{figure}

The pseudo-percolation threshold $R_{\text{perc}}$ is determined by requiring that a giant avalanche emerges when the average cluster radius $R_1$ reaches $R$ or close to $R$. We impliment it by requiring $R_1^{\text{perc}}=hR$, where $h\le 1$ is a proportionality constant close to unity. This condition yields
\begin{equation}
    f(R_1^{\text{perc}})=1-e^{-h}.
\end{equation}
For example, when $h=\frac{1}{2}$, then $R_1^{\text{perc}}\approx 10$, which is consistent with our simulation results. Thus, at this scale, a pseudo-percolation transition occurs. Beyond this threshold, however, the PP exhibits new features. Substituting Eq.~\ref{Eq:R1} into the scaling form gives
\begin{equation}
    \begin{split}
        P(\text{giant avalanche}) & \propto (L/R)^{-\gamma_1} R_1^{-\gamma_2} \\
        &= L^{-\gamma_1} R^{\gamma_1-\gamma_2} \left(-\ln[1-f(R)]\right)^{-\gamma_2},
    \end{split}
    \label{Eq:giantP_f_R}
\end{equation}
with $\bar{q}(R) > \bar{\mu}(R)$ for all $R$. This function is shown in Fig.~\ref{fig:f_R}. For all $L$, the fitting form
\begin{equation}
    f(R) = AR \exp\left[-\left(\frac{R}{R_0}\right)^{\xi}\right],
\end{equation}
provides good agreement with simulations, where $A$ depends on $L$ (e.g., $A\approx 9/2$ for $L=64$), while $R_0 \approx 1/3\times 10^{+2}$ and $\xi=0.2\pm 0.02$ are nearly independent of $L$. Since $f(R)\ll 1$ for all relevant $R$, we approximate $R_1 \approx R f(R)$, obtaining
\begin{equation}
    P(\text{giant avalanche}) \propto L^{-\gamma_1} R^{\gamma_1 - 2\gamma_2} \exp\left[\left(\frac{R}{R_0}\right)^{\xi}\right].
    \label{Eq:MFGiant}
\end{equation}
This predicts power-law growth for $R \lesssim 10$ (when $\gamma_1>2\gamma_2$), in agreement with simulations, while for $R>10$ it suggests exponential growth, which is not observed. Instead, the variance of $f(R)$ (Fig.~\ref{fig:f_R}, inset) grows sharply beyond $R\approx 10$, indicating strong fluctuations that invalidate the mean-field (MF) picture. Indeed, in Fig.~\ref{fig:pp} the PP displays some other non-trivial behaviors for $R\gtrsim 10$. This behavior is consistent with Eq.~\ref{Eq:R1}, which shows that $R_1$ decreases after the pseudo-percolation transition, compressing avalanches into smaller effective radii. As $R$ increases further, even a single unit circle may act as a \textit{large enough block} to contain sub-avalanches. In this case, Eq.~\ref{Eq:giantP_f_R} gives a next hierarchy of pseudo-percolation transition, which is obtained by replacing $L\to R$ and $R\to R_1$ and $R_1\to R_2$, according to which
\begin{equation}
    P_2(\text{giant avalanche}) = \left(\frac{L}{R}\right)^{-\gamma_1}\left[\left(\frac{R}{R_1}\right)^{-\gamma_1}R_2^{-\gamma_2}\right],
\end{equation}
where
\begin{equation}
    R_2 = -R_1 \ln[1-f(R_1)].
    \label{Eq:R2}
\end{equation}
Then, a second level pseudo-percolation transition arises when $R_2 \approx hR_1$ as described above. Iterating this logic leads to a hierarchy of transitions at scales $R_n$, given by
\begin{equation}
    \begin{split}
        & P_n(\text{giant avalanche}) = \left(\frac{R_{n-2}}{R_{n-1}}\right)^{-\gamma_1}R_{n}^{-\gamma_2}, \\
        & R_n = -R_{n-1}\ln[1-f(R_{n-1})].
        \label{Eq:HierarchicalEq}
    \end{split}
\end{equation}
This hierarchical structure is schematically illustrated in Fig.~\ref{fig:Hierarchical}. Each $n$th-level pseudo-percolation transition corresponds to percolation within the $n$th-level block. This hierarchy is the source of the same intermittent behaviors for the other exponents to be reported in the following sections.

\section{Numerical results}\label{SEC:sim}
In this section we preset the numerical results for both the local and global observables. For all figures plotted as a function of $R$, the values considered are $R=1, 2, 3, 4, 5, 6, 7, 8, 9, 10, 12, 14, 16, 20, 25,$ and $30$, arranged from top to bottom, respectively. In addition, the coefficient of determination ($R^2$) for the corresponding fittings is provided as an inset in the relevant figures.

To implement the sandpile toppling dynamics, we used Python in combination with several specialized libraries. 
The stack-based update procedure was implemented directly with \texttt{NumPy} arrays to ensure computational efficiency, and boolean masks (\texttt{toppled}, \texttt{listed}) were employed to monitor unstable sites and prevent redundant updates. For large-scale data management, including the storage of avalanche toppling matrices, we relied on the input/output utilities 
provided by \texttt{NumPy}.\\

As we will see in the following sections, the proposed hierarchical pseudo-percolation transitions give rise to scaling exponents with an intermittent behavior in terms of $R$, in a way much similar to that for PP (Fig.~\ref{fig:pp}). In this sense, we propose that the system \textit{displays crossover behavior among distinct fixed points}, each potentially associated with different scaling regimes. Within each crossover region, the measured exponents are consistent with those expected from the universality class of the corresponding fixed point.

\subsection{Local Observables}
Here we focus on the local observables $s$, $m$, and $d$, which characterize the local properties of avalanches. 
\begin{figure}
	\centering
	\includegraphics{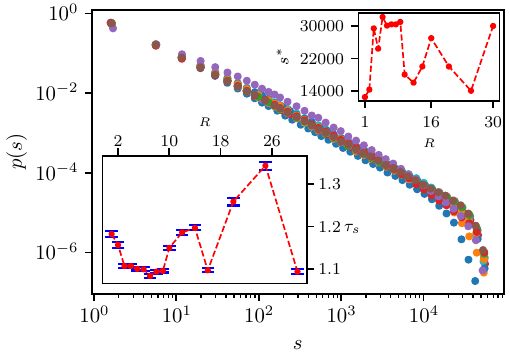}
	\caption{Avalanche size distributions for $L=256$ and varus $R$ values. They follow a power-law up to a cutoff $s^*$ (upper inset), which peaks near pseudo-percolation transitions. The size exponent $\tau_s$ ( is shown in lower inset as a function of $R$.}

	\label{fig:size}	
\end{figure}
Figure~\ref{fig:size} displays the probability distribution of avalanche sizes. The distributions follow a power-law form up to a characteristic cutoff scale, beyond which finite-size effects dominate, causing a rapid decay. This cutoff scale, denoted by $s^*$ (upper inset of Fig.~\ref{fig:size}), varies systematically with $R$. Interestingly, it reaches its maximum near the pseudo-percolation transition points, indicating that large-scale avalanches are favored in these regimes. For small $R$ (in particular $R=1$), the size exponent $\tau_s$ is close to the well-known BTW value, ${\tau_s^{\text{BTW}}}_{L\to\infty}\approx 1.21$~\cite{dhar1999some,lubeck1997large,najafi2021some}. As $R$ increases, $\tau_s$ gradually decreases and stabilizes around $1.10\pm 0.01$. Beyond $R\approx 10$, however, the exponent rises again, peaking at $R=15\sim 18$, close to where the system undergoes the first pseudo-percolation transition (lower inset of Fig.~\ref{fig:size}). It experiences its second peak at $R\sim 25$ where $\tau_s^{\text{max}}\approx 1.34\pm 0.01$ close to where the system undergoes its second pseudo-percolation transition. This behavior confirms that the critical exponents are not universal across scales, but rather evolve with $R$.  

\begin{figure}
	\centering
	\includegraphics{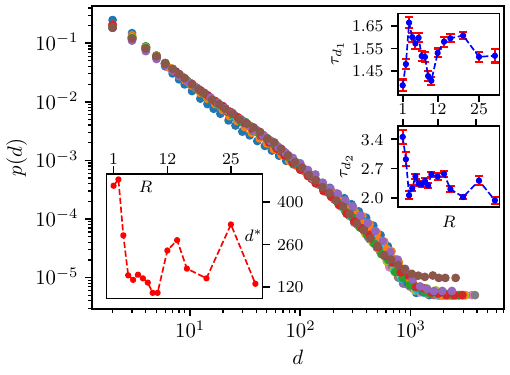}
	\caption{Avalanche duration distributions for $L=256$ and and varus $R$ values, showing power-law scaling with cutoff $d^*$ (lower inset). Two exponents, $\tau_{d1}$ and $\tau_{d2}$ are shown in terms of $R$ in the upper insets.}
	\label{fig:duration}	
\end{figure}

A similar non-universal behavior is observed for avalanche durations, as shown in Fig.~\ref{fig:duration}. 
The duration distributions follow a power-law with $R$-dependent exponents, 
which are modified near the pseudo-percolation points. 
As in the case of avalanche sizes, the distributions exhibit a cutoff scale, denoted by $d^*$ (lower inset of Fig.~\ref{fig:duration}), 
beyond which finite-size effects cause a rapid decay. 
The cutoff scale $d^*$ varies systematically with $R$ in the similar way as $s^*$, indicating that avalanche lifetimes are also sensitive to the crossover between fixed points. 
Moreover, the graphs show two distinct slopes in the power-law regime before the finite size effects appear. Consequently, two distinct exponents, $\tau_{d1}$ and $\tau_{d2}$, are distinguishable from the power-law regime. These exponents are shown in the upper insets of Fig.~\ref{fig:duration}, which confirm that $\tau_{d1}$ and $\tau_{d2}$ exhibit modifications in the vicinity of the pseudo-percolation points. In is notable that, such a \textit{bi-fractality} is also observed for $s$ for some especial $R$ values, but not for all, leading us to export the average slope for that case (lower inset of Fig.~\ref{fig:size}). In the cases where there are two or more fixed points of the dynamics of the system, such bi-fractality is a finite size effect, and disappear as $L\to\infty$ which is a signature of the crossover phenomena~\cite{najafi2012avalanche,najafi2016water,najafi2016bak,najafi2018sandpile,najafi2021some}.\\

\begin{figure}
	\centering
	\includegraphics{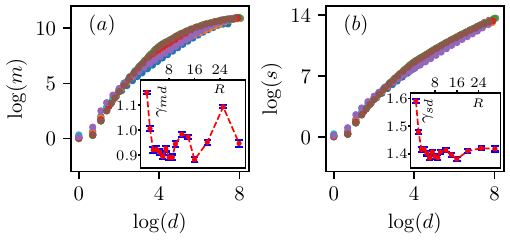}
	\caption{Scaling relations between avalanche mass $m$, size $s$, and duration $d$ for $L=256$ and varus $R$ values. The insets show the slopes in terms of $R$ which are the exponents $\gamma_{md}$ and $\gamma_{sd}$.}%
	\label{fig:m-d-s}	
\end{figure}

Finally, the scaling relations connecting avalanche size ($s$) and mass ($m$) with duration ($d$) are presented in Fig.~\ref{fig:m-d-s}. 
Both relations exhibit approximate power-law behaviors, in the correctly chosen intervals (at lease two decades). characterized by the exponents $\gamma_{sd}$ and $\gamma_{md}$. 
These exponents vary with $R$ and show noticeable changes near the pseudo-percolation points in a much similar way as $\tau_s$ and $\tau_d$ (For $\gamma_{md}$ it is clear, but for $\gamma_{sd}$ one should zoom in, since the data for $R=1$ is too large, so that the variance of the exponents is less evident). Taken together, these results suggest that the local observables are influenced by the hierarchical pseudo-percolation transitions: each transition appears to modify the effective scaling regime, leading to observable variations of the critical exponents with $R$. This indicates a possible connection between the global crossover behavior discussed earlier and the statistical properties of local avalanche observables.

\section{Global Geometric Structure}\label{SEC:fractal}
\begin{figure}
	\centering
	\includegraphics[width=0.9\linewidth]{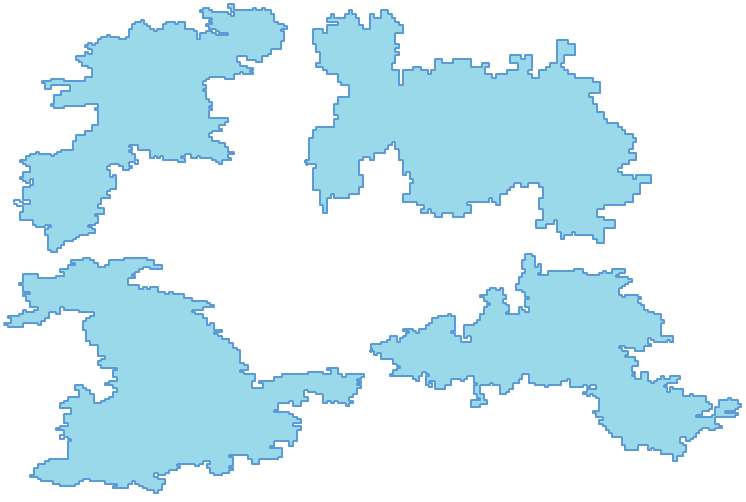}
	\caption{Representative avalanche samples for $R=5$. The shaded regions highlight toppled areas, illustrating that avalanches are generally not simply connected.}%
	\label{fig:Samples}	
\end{figure}

In addition to local observables, systems with self-similarity also exhibit global geometric features, which have been widely studied through both analytical approaches and numerical simulations. Such global properties often reveal structural aspects of the dynamics that remain hidden in local analyses. In this section, we focus on the global properties, especially the fractal characteristics of avalanches, and the geometry of their external perimeters with a hope that it provides further insight into the structural aspects of the model. 

As discussed previously, any avalanche consists of several connected sub-clusters, or sub-avalanches. To analyze the geometric properties of these structures, we first identify sub-avalanches using the Hoshen-Kopelman algorithm~\cite{hoshen1976percolation}, which assigns distinct labels to each connected component of a cluster. Some representative configurations are shown in Fig.~\ref{fig:Samples}.  

Each sub-avalanche has an external perimeter of length $l$ and a corresponding gyration radius $r$. Denoting the loop trace as $\{\mathbf{r}_i\equiv (x_i,y_i)\}_{i=1}^l$, the gyration radius is defined by
\begin{equation}
	\begin{split}
		&r^2 = \frac{1}{l} \sum_{i=1}^l \left| \mathbf{r}_i - \mathbf{r}_{\text{com}} \right|^2, \ \mathbf{r}_{\text{com}} = \frac{1}{l} \sum_{i=1}^l \mathbf{r}_i,
	\end{split}
\end{equation}
where $\mathbf{r}_{\text{com}}$ is the center of mass of the loop. The primary scaling exponent of interest is the fractal dimension $\gamma_{lr}=D_f$, defined through the relation
\begin{equation}
	\left\langle \log l \right\rangle = D_f \left\langle \log r \right\rangle + \text{const.},
\end{equation}
where $\langle \cdot \rangle$ denotes an ensemble average. According to the hyperscaling relation in Eq.~\ref{Eq:gamma}, the fractal dimension can also be expressed as
\begin{equation}
	\gamma_{lr}= D_f = \frac{\tau_r - 1}{\tau_l - 1}.
\end{equation}

\begin{figure}
    \centering \includegraphics{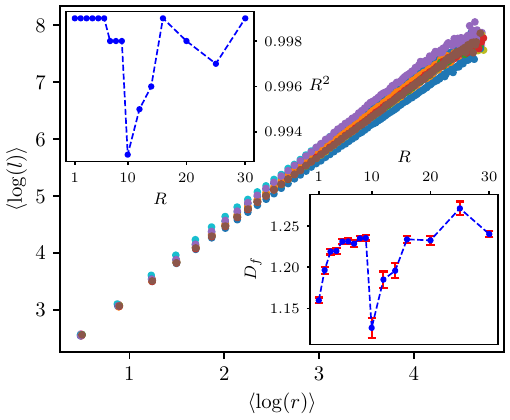}
    \caption{Fractal scaling of $\langle \log l \rangle$ vs $\langle \log r \rangle$ for $L=256$, showing linear behavior for and varus $R$ values (main figure). Insets display the fractal dimension ($D_f$) and the coefficient of determination ($R^2$) as functions of $R$.}

    \label{fig:fractal_dimension}
\end{figure}

Figure~\ref{fig:fractal_dimension} shows $\langle \log l \rangle$ versus $\langle \log r \rangle$, which display the expected linear behavior, the slope of which is $D_f$, along with the $R^2$ factor (not to be confused with the interaction range) in the insets. We observe that the fit quality is excellent for both small and sufficiently large $R$ values, while it deteriorates around $R\approx 10$, reflecting the crossover phenomena.  
A quantitative similar pattern is observed for $D_f$: $D_f(R=1)=1.16\pm0.01$, slightly below the BTW value $D_f^{\text{BTW}}=5/4$. It then saturates at $D_f(2\lesssim R\lesssim 8)=1.23\pm0.01$, shows a local minimum at $D_f(R=10)=1.13\pm0.01$, and gradually increases for $10\lesssim R\lesssim 25$, reaching a local maximum at $D_f(R=25)=1.27\pm0.01$. We see that the fractal dimension shows non-trivial behaviors around the Pseudo-percolation transition points. While we are not sure about the universal behavior of the model at these point, we just prefer to add that the dimension of the shortest paths of the critical percolation model is $d_{sp}=1.1307\pm 0.0004$~\cite{herrmann1988fractal}.

\begin{figure*}
	\centering
	\includegraphics{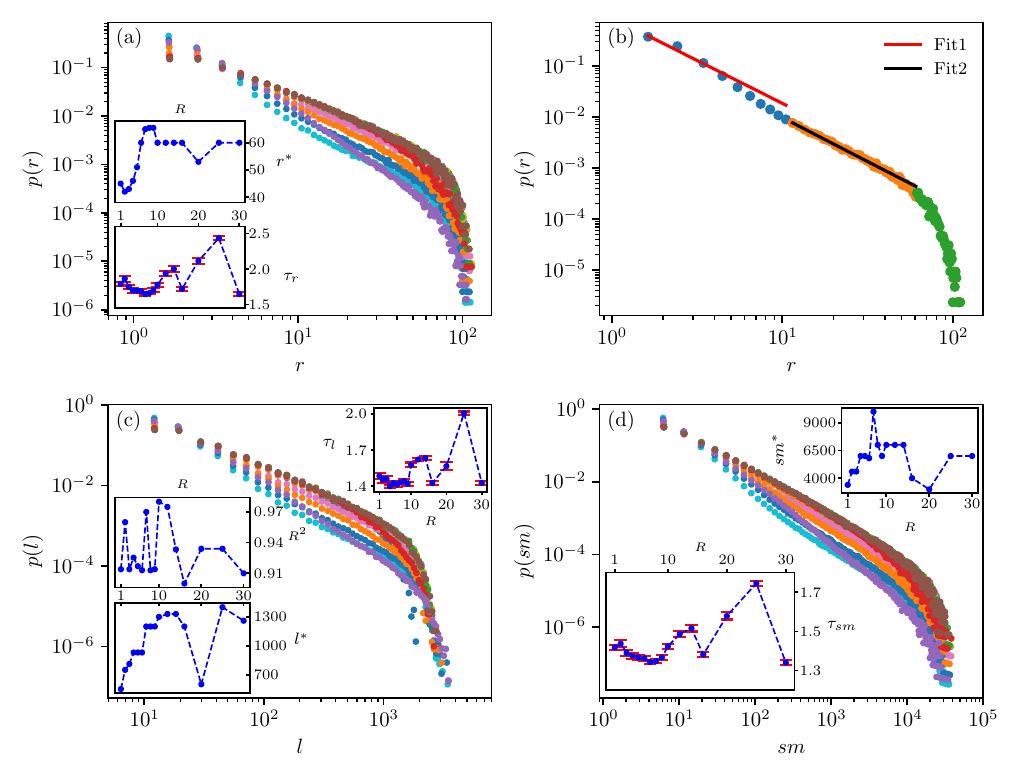}
	\caption{Distribution function of the (a) gyration radius $r$, (c) loop length $l$ and (d) sub-mass $sm$ for the sub-avalanches for $L=256$ and and varus $R$ values. The cutoff values ($r^*$, $l^*$ and $sm^*$), the $R^2$ of the fittings, and the exponents $\tau_r(R)$, $\tau_l(R)$ and $\tau_{sm}(R)$ are shown in the insets respectively. (b) Example illustrating distinct scaling regimes of $r$ for $R=12$.}

	\label{fig:gyration}
\end{figure*}

We also examined the distribution functions of the key geometrical observables, namely the gyration radius $r$, loop length $l$, and sub-avalanche mass $sm$. Among these, the gyration radius $r$ most directly reflects the spatial scaling behavior of the system. Figure~\ref{fig:gyration}a presents the distribution function of $r$ for different values of $R$. 
The distributions exhibit a cutoff scale, denoted by $r^*$ (upper inset of Fig.~\ref{fig:gyration}a), 
beyond which finite-size effects cause a rapid decay. 
The corresponding exponent $\tau_r$ exhibits a structure consistent with the hierarchical pseudo-percolation scenario. 
Specifically, starting from $\tau_{r}(R=1) = 1.5 \pm 0.1$, the value remains nearly constant with increasing $R$, 
from $\tau_{r}(R=10) = 1.78 \pm 0.03$, and then begins to increase. 
It subsequently decreases to a local minimum at $R = 16$, and reaches another maximum at $\tau_{r}(R=25) = 2.43 \pm 0.03$. 
These variations indicate that the exponents are distinct in the vicinity of transition points.
It is worth mentioning that, as in the previous cases, in some $R$ values, the power-law distributions display two distinct exponents (bi-fractality). A representative example is shown in Fig.~\ref{fig:gyration}b for $R=12$, where small scales follow $\tau_r^{\text{small scales}}\approx 2.1\pm0.1$, while large scales follow $\tau_r^{\text{large scales}}\approx 1.87\pm0.1$. This bi-fractality as in the above cases~\cite{najafi2012avalanche} is due to the fact that the properties of the model for the scales smaller than the correlation length (UV fixed-point behavior), and larger scales (corresponding to IR fixed-point) are different~\cite{goldenfeld2018lectures}. In Fig.~\ref{fig:gyration}a we ignored this complexity here, and reported the average slope here.

Figures~\ref{fig:gyration}c and~\ref{fig:gyration}d show the distribution functions for the loop length $l$ and sub-mass $sm$, respectively. 
Both quantities exhibit trends similar to the gyration radius, including features associated with hierarchical pseudo-percolation transitions. 
In Fig.~\ref{fig:gyration}c, one of the inset figures presents the coefficient of determination $R^2$ as a function of $R$, serving as an indicator of the quality of the scaling fit. 
Additional insets display the cutoff scale $l^*$ as a function of $R$, as well as the corresponding exponent $\tau_l(R)$. 
Similarly, in Fig.~\ref{fig:gyration}d, the upper inset shows the cutoff scale $sm^*$ versus $R$, while another inset reports the variation of the exponent $\tau_{sm}(R)$. 
These results collectively reinforce the consistency of the scaling behavior across different observables, further supporting the hierarchical pseudo-percolation scenario.

\section{Conclusion}

In this work, we proposed a new sandpile model with long-range interactions, regularized through a Yukawa cut-off ($R$) to avoid divergences that typically arise in critical systems with long-range interactions. This modification ensures a finite energy distribution across the lattice and enables a controlled study of collective avalanche dynamics. Our initial focus was on the percolation properties of the model, with special emphasis on the emergence of giant avalanches.

A detailed analysis of the avalanche percolation probability (PP) uncovered rich, scale-dependent behaviors (Fig.~\ref{fig:pp}). For small $R$, the PP grows monotonically with $R$, while at larger $R$ values we observed intermittent, successive pseudo-percolation phenomena—features not previously reported in sandpile models. To interpret these results, we developed a hierarchical mean-field (HMF) framework. In this picture, each disc at the $n$th hierarchical level consists of smaller sub-discs, and the global percolation dynamics arise from the nested pseudo-percolation probabilities within these substructures (Eq.~\ref{Eq:HierarchicalEq}).

We further examined both local and global observables as functions of $R$. These quantities generally exhibit power-law behavior, with critical exponents that vary systematically with $R$. The exponents were observed to undergo sharp changes near the pseudo-percolation transition points, reflecting the hierarchical nature of the system. The fractal dimension of the avalanche perimeter at the first transition point, matches the dimension of the shortest paths in critical percolation. Other exponents also shift significantly at the transition points. Our interpretation is that, as the system traverses the hierarchy of pseudo-percolation transitions, crossover takes place between different fixed points, which merit further investigation within the broader context of self-organized criticality.

In summary, our results show that incorporating a Yukawa cut-off into long-range sandpile models gives rise to hierarchical organization in terms of PP, which is accompanied with the substantial shifts in the critical exponents. These findings provide fresh perspectives on how long-range interactions shape self-organized critical phenomena.
  
\bibliography{refs}

\appendix

\end{document}